\documentclass[11pt]{article}
\usepackage{epsfig,amsfonts}
\def\f{\frac}

\def\nn{\nonumber}
\def\beq{\begin{equation}}
\def\eeq{\end{equation}}
\newcommand{\bea}{\begin{eqnarray}}
\newcommand{\eea}{\end{eqnarray}}
\newcommand{\bdm}{\begin{displaymath}}
\newcommand{\edm}{\end{displaymath}}
\long\def\symbolfootnote[#1]#2{\begingroup%
\def\thefootnote{\fnsymbol{footnote}}\footnote[#1]{#2}\endgroup}

\def\sq2{\sqrt{2}}

\def\msbar{\overline{\rm MS}}

\def\Bsg{B\to X_{\!s}\, \gamma}

\def\bratio{{\rm BR}(\Bsg)}

\def\Zbb{Z \to b \bar{b}}
\def\gaZbb{\Gamma (\Zbb)}

\newcommand{\smallz}{{\scriptscriptstyle Z}} 
\newcommand{\smallw}{{\scriptscriptstyle W}} %
\newcommand{\smallh}{{\scriptscriptstyle H^+}} %
\newcommand{\mz}{m_\smallz}
\newcommand{\mw}{m_\smallw}
\newcommand{\mh}{m_\smallh}

\def\as{\alpha_s}
\def\Au{A_u}
\def\muw{\mu_W}

\def\mt{\hat{m}_t}

\def\mb{\hat{m}_b}

\newenvironment{appendletterA}
 {
  \setcounter{section}{0}
  \setcounter{equation}{0}
  
 }{
 }

\begin{document}

\begin{titlepage}


{\flushright{
        \begin{minipage}{5cm}
          RM3-TH/10-02 \\
        \end{minipage}        }

}
\renewcommand{\thefootnote}{\fnsymbol{footnote}}
\vskip 2cm
\begin{center}
\boldmath
{\LARGE\bf QCD Corrections  in two-Higgs-doublet extensions \\[7pt] 
 of the Standard Model with Minimal Flavor Violation}\unboldmath
\vskip 1.cm
{\Large{G.~Degrassi$^{a}$ and P.~Slavich$^{b}$}}
\vspace*{8mm} \\
{\sl ${}^a$
    Dipartimento di Fisica, Universit\`a di Roma Tre and  INFN, Sezione di
    Roma Tre \\
    Via della Vasca Navale~84, I-00146 Rome, Italy}
\vspace*{2.5mm}\\
{\sl ${}^b$  LPTHE, 4, Place Jussieu, F-75252 Paris,  France}
\end{center}
\symbolfootnote[0]{{\tt e-mail:}}
\symbolfootnote[0]{{\tt degrassi@fis.uniroma3.it}}
\symbolfootnote[0]{{\tt slavich@lpthe.jussieu.fr}}

\vskip 0.7cm

\begin{abstract}
  We present the QCD corrections to $R_b$ and to the $\Delta B=1$
  effective Hamiltonian in models with a second Higgs field that
  couples to the quarks respecting the criterion of Minimal Flavor
  Violation, thus belonging either to the $({\bf 1},{\bf 2})_{1/2}$ or
  to the $({\bf 8},{\bf 2})_{1/2}$ representation of $SU(3)\times
  SU(2)\times U(1)$.  After the inclusion of the QCD corrections, the
  prediction for $R_b$ becomes practically insensitive to the choice
  of renormalization scheme for the top mass, which for the type-I and
  type-II models translates in a more robust lower bound on $\tan
  \beta$. The QCD-corrected determinations of $R_b$ and $\bratio$ are
  used to discuss the constraints on the couplings of a (colored)
  charged Higgs boson to top and bottom quarks.
\end{abstract}
\vfill
\end{titlepage}    
\setcounter{footnote}{0}


\section{Introduction}

One of the main goals of the present experimental program at the
Tevatron and at the Large Hadron Collider (LHC) is the search for the
Higgs boson(s) in order to elucidate the mechanism of electroweak
symmetry breaking (EWSB). In the Standard Model (SM), the latter is
realized in the most economical way via a single Higgs doublet.  This
minimal realization predicts a single neutral Higgs boson, whose mass
can be constrained, from electroweak precision data and the direct
search limit from LEP, to be lighter than $\sim 200$ GeV. However, at
the moment, there is no direct experimental evidence for a neutral
Higgs boson or any other scalar particle like, for example, a charged
boson that can be present in models with a nonminimal Higgs sector.

The LHC is going to explore physics up to the TeV scale in order to
search for the Higgs boson, as well as for any new phenomenon that
would confirm the widespread expectation that the picture of particle
physics in terms of the SM is incomplete.  However, new particles with
mass in the TeV range that couple to quarks at the tree level can
modify the predictions for Flavor Changing Neutral Current (FCNC)
processes. Thus any extension of the SM, starting from the simplest we
can think of, a Two-Higgs-Doublet Model (2HDM), needs to face the
problem of avoiding conflicts with the strict limits on FCNC
processes.
 
Glashow and Weinberg addressed this issue proposing the principle of
Natural Flavor Conservation (NFC) \cite{GW}, which requires that the
matrices of Yukawa couplings to up and down quarks for all the Higgs
fields be diagonal in the basis where the quark mass matrices,
$M^{U,D}$, are diagonal. This implies that, with the exception of models 
with vectorlike quarks that mix with the ordinary ones, NFC models do
not have tree-level FCNC couplings. In the 2HDM case, NFC can be realized
imposing the sufficient condition that each of the quark mass matrices
is obtained from a single Higgs field. This can be enforced via a
${\cal Z}_2$ symmetry that acts differently on the two Higgs doublets,
leading to two possibilities usually referred to as type-I (i.e., the
model in which both up and down quarks get their masses from Yukawa
couplings to the same Higgs doublet) and type-II models (where up and
down quarks get their masses from Yukawa couplings to different Higgs
doublets).

A less-restrictive way to suppress FCNC processes, still avoiding
conflict with the experimental bounds, is to consider the criterion of
Minimal Flavor Violation (MFV) \cite{DGIS}, which amounts to assuming
that all the new flavor-changing transitions, including those mediated
at the tree level by electrically neutral particles, are controlled by
the Cabibbo-Kobayashi-Maskawa (CKM) matrix. Thus, the MFV hypothesis
requires that all the flavor-violating interactions of the new
particles present at the TeV scale be linked to the known structure of
the Yukawa couplings.

The enforcement of the MFV hypothesis to the case of multi-Higgs
models has been recently investigated by several groups
\cite{MW,BBR,PT}. In particular, in ref.~\cite{MW} it has been shown,
via group-theoretic arguments, that the MFV hypothesis can be enforced
requiring that all the Higgs Yukawa-coupling matrices be composed from
the pair of matrices $Y^U$ and $Y^D$ that are responsible for the
breaking of the $SU(3)_{Q_{L}} \times SU(3)_{U_{R}} \times
SU(3)_{U_{D}}$ quark flavor symmetry. This requirement restricts the
allowed $SU(3)\times SU(2)\times U(1)$ representations of the Higgs
fields that can couple to the quarks to either be equal to that of the
SM Higgs field, i.e.~$({\bf 1},{\bf 2})_{1/2}$, or transform as $({\bf
  8},{\bf 2})_{1/2}$. Examples of the former case, besides the NFC
type I and II models, are the aligned model of ref.~\cite{PT} or the
class of 2HDM presented in ref.~\cite{BGL}. The latter case is quite
different, because the second field does not acquire a vacuum
expectation value (vev) and does not mix with the SM Higgs field. Thus
the scalar spectrum of this model contains a CP-even, color-singlet
Higgs boson (the usual SM one) and three color-octet particles, one
CP-even, one CP-odd and one electrically charged, which are split in
mass proportionally to the SM-Higgs vev \cite{MW}. These colored
scalar particles give rise to an interesting phenomenology for the
LHC, not only because -- if they are not too heavy -- they can be
directly produced, but also because their indirect effects can
influence flavor, electroweak and Higgs physics
\cite{MW,GrW,IKM,BDV,BTZ}.

Models with a second Higgs doublet present a new and interesting
phenomenology, in particular related to the presence of a charged
scalar. In the flavor sector, decays mediated by a weak charged
current are the natural place where effects due to a charged Higgs
boson, $H^+$, can show up. In the electroweak sector the observable
$R_b \equiv \gaZbb/\Gamma(Z \to$ {\em hadrons}) shows a sensitivity to
$H^+$ because of the specific vertex corrections introduced by the
interaction of $H^+$ with the top and bottom quarks.  Many studies
(for the most recent see, e.g., refs.~\cite{Gfit,CKMfit,UTfit}) used
various combinations of flavor and electroweak observables to
constrain the parameter space of the type-II 2HDM, which garnered most
of the attention because of its property of having the same
Higgs-sector realization as the Minimal Supersymmetric Standard Model
(MSSM). Other studies \cite{Ida,Moretti,MS} explored the parameter
space of 2HDMs unconstrained by a ${\cal Z}_2$ symmetry, with the
second Higgs doublet still in the $({\bf 1},{\bf 2})_{1/2}$
representation.

The theoretical accuracy of the predictions in the 2HDM with MFV is
not yet at the same level as in the SM. Here we take a first step in
improving this situation, by (re)considering the QCD corrections to
two observables, $R_b $ and $\bratio$, which allow to set important
constraints on the mass and couplings of the charged scalar. The QCD
corrections can play a relevant role in reducing the error of the
theoretical predictions, a well-known example of this fact being
indeed the radiative decay of the $B$ meson.  The present knowledge in
the 2HDMs of the two observables we are considering can be summarized
in this way: in the case of models with a second Higgs doublet in the
$({\bf 1},{\bf 2})_{1/2}$ representation, the complete one-loop
calculation of $\gaZbb$ is available \cite{DGHK}, but (to our
knowledge) no QCD corrections to the charged-scalar contributions are
known.  The process $\Bsg$ is instead fully known at the
Next-to-Leading Order (NLO) in QCD \cite{CMM,CDGG1,CRS,BG}. The case
with colored scalars in the adjoint representation of $SU(3)$ is less
studied.  The one-loop charged-scalar contribution to $\gaZbb$ was
reported in ref.~\cite{GrW} (see also ref.~\cite{HL}) while for the
radiative decay of the B meson only a partial result for the Leading
Order (LO) Wilson coefficients of the magnetic and chromo-magnetic
operators has been presented \cite{MW}.

In this paper we present the QCD corrections to the contribution to
$\gaZbb$ of a charged scalar in either the $({\bf 1},{\bf 2})_{1/2}$
or the $({\bf 8},{\bf 2})_{1/2}$ representation. Concerning $\Bsg$, we
compute the ${\cal O}(\as)$ contribution to the Wilson coefficients
due to a colored charged scalar in the $({\bf 8},{\bf 2})_{1/2}$
representation.  This is the missing piece to achieve NLO predictions
for $\bratio$ for all 2HDMs with MFV. Because of the specific
interactions of the colored scalar with the gluons, the Wilson
coefficients cannot be simply obtained by an appropriate color-factor
rescaling of the known $({\bf 1},{\bf 2})_{1/2}$ result.

The paper is organized as follows: in the next section we discuss the
couplings of the charged Higgs boson in the different realizations of
the 2HDM with MFV. In section 3 we present the results for the
QCD-corrected contribution to $\gaZbb$ due to a charged scalar,
covering both cases of color-singlet and color-octet particle. We show
that, after the inclusion of the QCD correction, the prediction for
$R_b$ is practically insensitive to choice of an $\msbar$ or on-shell
(OS) renormalization scheme for the top mass. The bounds set by $R_b$
on the $tbH^+$ coupling are also shown.  Section 4 contains the result
for the NLO Wilson coefficients in the $\Delta B =1$ effective
Hamiltonian (the explicit analytic expressions are presented in the
appendix).  The known results for the colorless 2HDMs are recovered,
while the case of a colored charged scalar is fully new.  The
restrictions imposed by $\Bsg$ on the charged-scalar interaction with
quarks are discussed.  Finally, in section 5 we present our
conclusions.


\section{Minimally flavor violating 2HDMs}

In a generic 2HDM it is always possible to rotate the two Higgs fields
to a basis in which only one of them, which we denote as $\Phi_1$,
gets a vev~\cite{habdav}. In this basis, we write the Yukawa
interactions of the Higgs fields with the quarks as
\beq
\label{lagYuk}
-{\cal L}_Y ~=~
\bar{q}_L\, \widetilde{\Phi}_1\, Y^U u_R ~+~  \bar{q}_L\, \Phi_1\, Y^D d_R
~+~\bar{q}_L \,\widetilde{\Phi}^{(a)}_2\, T_R^{(a)}\,\bar{Y}^U u_R ~+~  
\bar{q}_L \,\Phi_2^{(a)}\, T_R^{(a)}\, \bar{Y}^D d_R~~+~{\rm h.c.}~,
\eeq
where $\widetilde{\Phi}_i \equiv i\sigma_2\Phi_i^*$, and the Yukawa
couplings $Y^{U,D}$ are $3\!\times\!3$ matrices in flavor space such
that $M^{U,D} = Y^{U,D}\,\langle \Phi_1^0\rangle$. The possibility of
a colored second Higgs doublet is encoded in the matrices $T_R^{(a)}$
that act on the quark fields. In the usual colorless 2HDM $T_R$ is
equal to the identity matrix in color space. On the other hand, for a
colored Higgs doublet in the adjoint representation of $SU(3)$
$T_R^a=T_F^a\: (a=1,8)$, the matrices of the fundamental
representation.  The MFV condition amounts to requiring that the
Yukawa coupling matrices of the second doublet, $\bar Y^{U,D}$, be
composed of combinations of the matrices $Y^{U,D}$, and transform
under the $SU(3)_{Q_{L}} \times SU(3)_{U_{R}} \times SU(3)_{U_{D}}$
quark flavor symmetry in the same way as $Y^{U,D}$ themselves. We can
decompose the matrices $\bar Y^{U,D}$ as
\beq
\label{MFVdeco}
\bar Y^U = A_u \,\left( 1 + \epsilon_u\,Y^U Y^{U\,\dagger} 
+ \ldots\right)\,Y^U~,~~~~~
\bar Y^D = A_d \,\left( 1 + \epsilon_d\,Y^U Y^{U\,\dagger} 
+ \ldots\right)\,Y^D~,
\eeq
where in principle $A_{u,d}$ and $\epsilon_{u,d}$ are arbitrary
complex coefficients. The ellipses in eq.~(\ref{MFVdeco}) denote terms
involving powers of $Y^DY^{D\,\dagger}$ as well as terms involving
higher powers of $Y^UY^{U\,\dagger}$. In the following, we will assume
that the only significant deviations from proportionality between
$\bar Y^{U,D}$ and $Y^{U,D}$ are controlled by the Yukawa coupling of
the top quark, and that terms involving higher powers of the Yukawa
matrices are suppressed (e.g., because they are generated at higher
loops). If we further require that there are no new sources of CP
violation apart from the complex phase in the CKM matrix, the
coefficients $A_{u,d}$ and $\epsilon_{u,d}$ must be real.  Finally,
the case $\epsilon_u=\epsilon_d=0$ corresponds to the NFC situation in
which the Yukawa matrices of both Higgs doublets are aligned in flavor
space.

The processes $\Zbb$ and $\Bsg$ that we will consider in sections 3
and 4 involve loops with a charged Higgs boson and a top quark. Under
the assumptions implicit in eq.~(\ref{MFVdeco}), the interaction
between the quarks and $H^+$ is controlled by the Lagrangian
\beq
\label{lagH}
{\cal L}_{H^+}= - \frac{g}{\sqrt{2\,} \mw}\, 
\sum_{i,j=1}^3\,\bar u_i \,T_R^{(a)}
\left( A^i_u \,m_{u_i}\,\frac{1-\gamma_5}{2}-
A^i_d\, m_{d_j}\,\frac{1+\gamma_5}{2}\right)\, V_{ij}\,d_j\, 
H^+_{(a)} +{\rm h.c.}~,
\eeq
where $g$ is the $SU(2)$ coupling constant, $i,j$ are generation
indices, $m_{u,d}$ are quark masses, $V$ is the CKM matrix. The
family-dependent couplings $A_{u,d}^i$ read
\beq
\label{shift}
A_{u,d}^i = A_{u,d}\,\left(1 +
  \epsilon_{u,d}\,\frac{m_t^2}{v^2}\,\delta_{i3}\right)~, 
\eeq
where $v = \langle \Phi_1^0 \rangle$. It appears from
eq.~(\ref{shift}) that, when we neglect the masses of the light
quarks, the effect on the charged-Higgs couplings arising from the
$Y^UY^{U\,\dagger}$ terms in eq.~(\ref{MFVdeco}) is limited to a shift
in the couplings involving the top quark. Since those are the only
couplings that enter our computations, in sections 3 and 4 we will
drop the family index from $A_{u,d}^i$ without ambiguity.

The term $\epsilon_d Y^UY^{U\,\dagger}$ entering the expression for
$\bar Y^D$ in eq.~(\ref{MFVdeco}) also induces a flavor-changing
interaction with the down quarks for the neutral component of the
second Higgs doublet. However, this interaction does not affect the
computation of $\gaZbb$, and its contribution to $\Bsg$ is negligible
with respect to the charged-Higgs contribution as long as
$\epsilon_d\,A_d/A_u \ll (v/m_b)^2$. Other FCNC processes such as
$B\bar B$ mixing would put bounds on the combination
$\epsilon_d\,A_d$, but this will not be relevant to the discussion
that follows.

In the notation of eq.~(\ref{lagH}), the type-I and type-II models are
specified by $T_R^a$ equal to the identity matrix in color space, and
by the real (and family-universal) coefficients
\bea
A_u^i =  A_d^i &=& 1/\tan \beta ~~~~~~ \mbox{(type I)}, \\
A_u^i=-1/A_d^i &=& 1/\tan \beta ~~~~~~ \mbox{(type II)},
\eea
where $\tan \beta$ is the ratio of the vevs of the two Higgs doublets
in the basis where each of the quark mass matrices is obtained from a
single Higgs field.

According to our discussion, the MFV hypothesis includes two other
possibilities, namely color-singlet and color-octet Higgs doublet that
couple to the quarks with arbitrary coefficients\footnote{Even models
  with generic Yukawa matrices not satisfying the MFV hypothesis show a
  structure of couplings with arbitrary coefficients $A_{u,d}^i$.  To
  be phenomenologically viable, the dangerous FCNC effects should be
  sufficiently suppressed via some specific assumption like, e.g., a
  specific texture of the Yukawa matrices \cite{CS}.} $A_{u,d}^i$.  We
are going to refer to the first category (singlet) as type-III model,
while the second possibility (octet) will be called type-C model.

\section{Charged-Higgs contribution to $R_b$ including QCD
  corrections}

We begin by discussing the radiatively corrected partial decay width
of the $Z$ boson in a quark-antiquark pair in a model with an
additional Higgs doublet.  We write it as
\beq
\Gamma(Z \to q \bar{q}) = N_c\, \frac{G_\mu}{\sqrt2}\, \frac{\mz^3}{3\, \pi}
 \,\left[ (\bar{g}^{q}_L)^2 + (\bar{g}^{q}_R)^2 \right] K_q~,
\label{eq:0}
\eeq
where $N_c$ is the color factor ($N_c=3$), $\bar{g}^{q}_{(L,R)}$ are
the left-handed and right-handed $Zq\bar q$ couplings, written in terms
of the radiative parameter $\rho_q$ and the radiatively corrected sine
of the Weinberg angle $\bar{s}_W^q$ as ($T_3^q$ is the third component
of the weak isospin, $Q_q$ is the electric charge in unit $e$)
\bea 
 \bar{g}^{q}_L &=& \sqrt{\rho_q} \,\left[ T_3^q 
- Q_q\, (\bar{s}_W^q)^2 \right]~,\nn \\
 \bar{g}^{q}_R &=& - \sqrt{\rho_q}\, Q_q (\bar{s}_W^q)^2~,
\eea
while the factor $K_q$ contains the QCD, QED and quark-mass
corrections. The latter has been computed up to ${\cal O}(\as^3)$ in
ref.~\cite{CKK}, and at the lowest order it reads:
\beq
 K_q = 1 + C_F \frac{3\as}{4 \pi} + Q_q^2\, \frac{3\alpha}{4 \pi} -
\frac34 \frac{\mu_q}{(\bar{g}^{q}_L)^2 + (\bar{g}^{q}_R)^2} + {\cal O}(\as^2)~,
\eeq 
where $\mu_q = m_q^2/\mz^2$ and $C_F =4/3$.

We assume that the oblique corrections due to the second doublet are
negligible, as happens when the spectrum of the additional states is
approximately custodially symmetric. Then the effect of the second
Higgs doublet is concentrated in the vertex corrections to
$\gaZbb$. Thus, defining
\bea
\rho_q &=& \rho_q^{SM} + \delta \rho_q~, \\
(\bar{s}_W^q)^2 &=&  (\bar{s}_{W}^q)^2_{SM} + \delta (\bar{s}_W^q)^2~,
\eea
we have $\delta \rho_{(q \neq b)} = \delta (\bar{s}_W^{(q \neq b)})^2
=0$. In the limit of neglecting the mass of the $Z$ boson with respect
to the masses of the top quark and the charged Higgs boson we find
\bea
\delta \rho_b \hspace{-0.15cm} &=&
\frac1{T_3^b}\,\frac{\alpha}{4 \pi s^2_W} \,C^1_R\,
\left[ \left( \frac{|\Au|\,\mt}{\sqrt{2} \mw}\right)^2 + 
\left(\frac{|A_d| \,\mb}{\sqrt{2} \mw}\right)^2 
\right] \left[  \,f_1 (t_h) + \frac{\as}{4 \pi} 
 \left(C_F \, f_2 (t_h) + C_R^2 \, f_3 (t_h) 
\right) \right]\,,~~~ \label{eq:1}\\\nn\\
\delta (\bar{s}_W^b)^2  \hspace{-0.15cm} &=& -\frac12 \delta \rho_b s^2_W + 
\frac1{2\,Q_b} \, \frac{\alpha}{4 \pi s^2_W} \, C^1_R 
\left(\frac{|A_d| \,\mb}{\sqrt{2} \mw}\right)^2 
\left[ f_1(t_h) + \frac{\as}{4 \pi} 
 \left(C_F  \,f_2 (t_h) + C_R^2 \, f_3 (t_h) 
\right) \right]~, 
\label{eq:2}
\eea
where $C^1_R=1, C^2_R = 0 \:[C^1_R=C_F, C^2_R = C_A=N_c]$ for Higgs
fields in the $({\bf 1},{\bf 2})_{1/2}\: [({\bf 8},{\bf 2})_{1/2}]$
representation, and we omit an overall factor $|V_{tb}|^2 \approx
1$. In eqs.~(\ref{eq:1}) and (\ref{eq:2}) $t_h= \mt^2/\mh^2$, where
$\hat{m}_q$ is the $\msbar$ quark mass at the scale $\mu$ and $\mh$ is
the OS $H^+$ mass. The explicit expressions for the functions
$f_i(x)$ are
\bea
f_1(x) & = & \frac{x}{x-1}-\frac{x \,\ln x}{(x-1)^2}~, \\
f_2(x) & =&
 -\frac{6 \,x\left(x-2 \right)}{(x-1)^2} {\rm Li}_2 \left(1-\frac1x\right)+
\frac{ x (-27 +11 x)}{(x-1)^2}+\frac{x (25-9 x) \ln x}{ (x-1)^3}\nn \\
 &+& 
\left( \frac{6x(3-x)}{(x-1)^2} - \frac{12 x \ln x}{(x-1)^3} \right)
\ln \frac{\mt^2}{\mu^2} -3 f_1(x)~,\\
f_3(x) & = &  \frac{3 
x }{ (x-1)} {\rm Li}_2\left(1-\frac1x\right)
+\frac{3 x \left(1-2 x + x^2+ \ln^2 x\right)}{ (x-1)^3}-
\frac{6 x \ln x}{(x-1)^2}~,
\eea
where the last term ($-3 f_1$) in the function $f_2$ is introduced to
avoid double counting due to the correction factor $K_q$ in
eq.~(\ref{eq:0}), and in the $tbH^+$ coupling we have also kept the
contribution proportional to the bottom mass\footnote{Terms
  proportional to $\mb$, relevant only for very large values of $A_d$,
  can also arise from vertices with neutral scalars.}.

The one-loop terms in $\delta \rho_b$ and $\delta (\bar{s}_W^b)^2$
agree with the results\footnote{In ref.~\cite{HL} there is a misprint
  in the overall normalization of the $\delta g^{L,R}$ couplings.} of
refs.~\cite{DGHK,HL}.  The two-loop terms were obtained following the
lines of the analogous SM calculation that was performed by several
groups, via different methods, in the early nineties \cite{Zbbme}. The
SM correction can be actually obtained by considering the SM
Lagrangian in the limit of vanishing gauge coupling constants, the
so-called gaugeless limit of the SM \cite{BBCCV}. In this limit the
gauge bosons play the role of external sources, and the propagating
fields are those of a Yukawa theory with massless Goldstone
bosons. Indeed, in the limit $\mh \to 0$, the ${\cal O}(\alpha \as
\mt^2/\mw^2)$ corrections in eq.~(\ref{eq:1}) agree with the known SM
result.

To express the corrections in terms of the OS top mass $m_t$, we
must expand the $\msbar$ top mass entering the one-loop part as
$\hat{m}_t = m_t + \delta m_t$, with
\beq
\delta m_t = \frac{\as}{4 \pi} \,C_F \left( 3 \ln 
\frac{m_t^2}{\mu^2} -4 \right) \, m_t~.
\eeq
For the terms proportional to $|A_u|^2$ and $|A_d|^2$ in
eqs.~(\ref{eq:1}) and (\ref{eq:2}), this amounts to making the
substitution $\mt \to m_t$ and replacing the function $f_2$ with the
OS counterparts $f_2^u$ and $f_2^d$, respectively:
\beq
f_2^{u}(x) ~=~
f_2(x) ~+~ \frac{8\pi}{\as\,C_F}\,\frac{\delta m_t}{m_t}\,
\left[f_1(x) + x\,\frac{\partial f_1(x)}{\partial x}\right]~,~~~~~~ 
f_2^{d}(x) ~=~
f_2(x) ~+~ \frac{8\pi}{\as\,C_F}\,\frac{\delta m_t}{m_t}\,x\,\frac{\partial f_1(x)}{\partial x}~.
\eeq
The explicit dependence on the renormalization scale cancels out in
the function $f^u_2$:
\beq
f_2^{u}(x)
~=~
 -\frac{6 \,x\left(x-2 \right)}{(x-1)^2} {\rm Li}_2 \left(1-\frac1x\right)+
\frac{ 3 x }{(x-1)}-\frac{9 x \ln x}{ (x-1)^2}-3 f_1(x)~,
\eeq
whereas $f^d_2$ has a residual dependence on $\mu$ which is
compensated for by the implicit scale dependence of the $\hat m_b^2$
entering the one-loop parts of eqs.~(\ref{eq:1}) and
(\ref{eq:2}). Using the OS bottom mass $m_b$ would remove this
residual scale dependence, but it would introduce large logarithms of
the ratio $m_t/m_b$ in the two-loop part of the corrections.

\begin{figure}[t]
\begin{center}
\epsfig{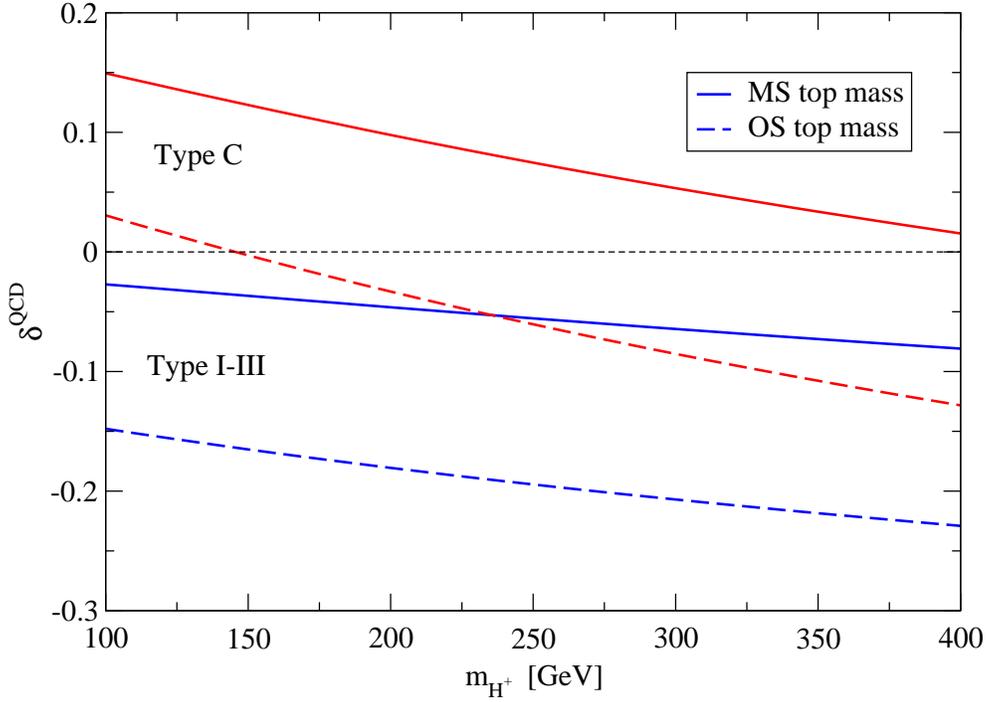}
\end{center}
\vspace{-2mm}
\caption{the ratio of two-loop to one-loop charged-Higgs contributions
  to the $Zb\bar b$ vertex as a function of $\mh$, with the top mass
  expressed in the $\msbar$ (solid lines) or OS (dashed lines)
  renormalization scheme. The upper (red) curves are for the model
  with color-octet Higgs (type C), while the lower (blue) curves are
  for the models with color-singlet Higgs (types I--III).}
\label{fig1}
\end{figure}

In fig.~\ref{fig1} we plot, as a function of $\mh$, the ratio
$\delta^{QCD}$ between the two-loop and one-loop contributions in the
terms proportional to $|A_u|^2$ in eq.~(\ref{eq:1}), for the two cases
of color-singlet (lower set of lines) and color-octet (upper set of
lines) charged Higgs boson. For each case, we show $\delta^{QCD}$ as
obtained using either the central value of the physical (OS) top mass,
$m_t = 173.1$ GeV \cite{top} (dashed lines), or the corresponding
$\msbar$ value $\mt(m_t)= 163.5$ GeV (solid lines), with the
appropriate formulae for the two-loop function $f_2$. It can be seen
from the figure that, for models of types I--III (i.e.~with
color-singlet charged Higgs) the two-loop corrections are always
negative, and they are substantially larger when the OS top mass is
used in the one-loop part than when the $\msbar$ mass is used. On the
other hand, for the model of type C (with color-octet charged Higgs)
there is an overall upward shift in the two-loop correction due to the
additional function $f_3$ in eq.~(\ref{eq:1}), with the result that
the sign and relative size of the corrections in the OS and $\msbar$
cases depend on the Higgs mass. For low values $\mh \approx 150$ GeV,
the two-loop correction approaches zero if the OS top mass is used,
and it is positive and relatively large if the $\msbar$ mass is
used. Conversely, for larger values $\mh \approx 400$ GeV the two-loop
correction approaches zero if the $\msbar$ top mass is used, and it is
negative and relatively large if the OS mass is used. As a result, we
will see that for 2HDMs of types I--III a reliable approximation of
the two-loop result for $R_b$ could be obtained by using the one-loop
result expressed in terms of the $\msbar$ top mass. On the other hand,
a precise determination of $R_b$ in the 2HDM of type C requires the
inclusion of the two-loop part of the $Zb\bar b$ vertex correction.

From eqs.~(\ref{eq:0})--(\ref{eq:2}) we can construct the observable
$R_b $, which can be written as
\beq
\label{defRb}
\frac1{R_b} = 1 + \frac{\sum_{(q \neq b)} 
 \left[ (\bar{g}^{q}_L)^2 + (\bar{g}^{q}_R)^2 \right] K_q}
{ \left[ (\bar{g}^{b}_L)^2 + (\bar{g}^{b}_R)^2 \right] K_b}
\equiv 1+ \frac{S_b}{s_b} C_{b}
\eeq
where
\beq
\label{defSb}
S_b = \sum_{(q \neq b)} s_q~;~~~~~~~
s_q = \left[ (\bar{g}^{q}_L)^2 + (\bar{g}^{q}_R)^2 \right]
      \left( 1 +  Q_q^2\, \frac{3\alpha}{4 \pi} \right)~.
\eeq
Using the results of ref.~\cite{LEPEWWG} we find
$S_b=0.6607$. Concerning the SM part of $s_b$, to avoid relying
indirectly on the measured value of $R_b$, we follow
ref.~\cite{CKMfit} and compute it using the values $\rho_b^{SM} =
0.99426$ \cite{LEPEWWG} and $(\bar{s}_W^b)^2_{SM} = (1.0063)\times
0.23153=0.23299$, the latter obtained from the measured value of
$\sin^2 \theta^{lept}_{eff}=0.23153 \pm 0.00016$ corrected for the
top-induced contributions specific to the $Zb\bar{b}$ vertex\footnote{
  In ref.~\cite{CKMfit} the correcting factor 1.0063 was not
  introduced.}.  Finally the factor $C_{b}$ that includes QCD and mass
corrections is obtained from ref.~\cite{CKK}. We find, for $\as =
0.118$, $C_b = 1.0086$.

With the values specified above for the various parameters entering
eqs.~(\ref{defRb}) and (\ref{defSb}) we find a SM prediction
$R_b^{\rm{\scriptscriptstyle SM}} = 0.21580$, nearly $1\sigma$ below
the measured value $R_b^{\rm exp} = 0.21629 \pm 0.00066$
\cite{LEPEWWG}. Since the charged-Higgs contributions to
$\bar{g}_{L}^b$ and $\bar{g}_{R}^b$ have the effect of further
lowering $R_b$, stringent bounds can be imposed on the parameters
$\mh$ and $A_u$ by the requirement that the predicted value of $R_b$
in a 2HDM be not too far from $R_b^{\rm exp}$. On the other hand,
$R_b$ has little sensitivity on $A_d$, because the terms in
eqs.~(\ref{eq:1}) and (\ref{eq:2}) controlled by the latter are
suppressed by $\hat m_b$. Therefore, for the models of types III and C
in which $A_d$ is a free parameter we will simplify our discussion by
setting $A_d=0$. In the 2HDMs of type I and II the parameter $A_d$ is
related to $A_u$ and cannot be set independently to zero. However, due
to the strong suppression of the contributions controlled by $A_d$, in
most of the parameter space the predictions of $R_b$ obtained in those
two models do not differ significantly from the predictions obtained
in the type-III 2HDM with $A_d=0$. More specifically, the predictions
of the type-I 2HDM, in which $A_d=A_u$, are virtually
indistinguishable from those of the type-III 2HDM with $A_d=0$ for all
the values of $A_u$ consistent with the measured value of $R_b^{\rm
  exp}$. In the 2HDM of type II, on the other hand, $A_d=-1/A_u$, and
the predictions of $R_b$ differ from the ones obtained in the type-III
2HDM with $A_d=0$ only for very small values of $A_u$.

\begin{figure}[p]
\begin{center}
\epsfig{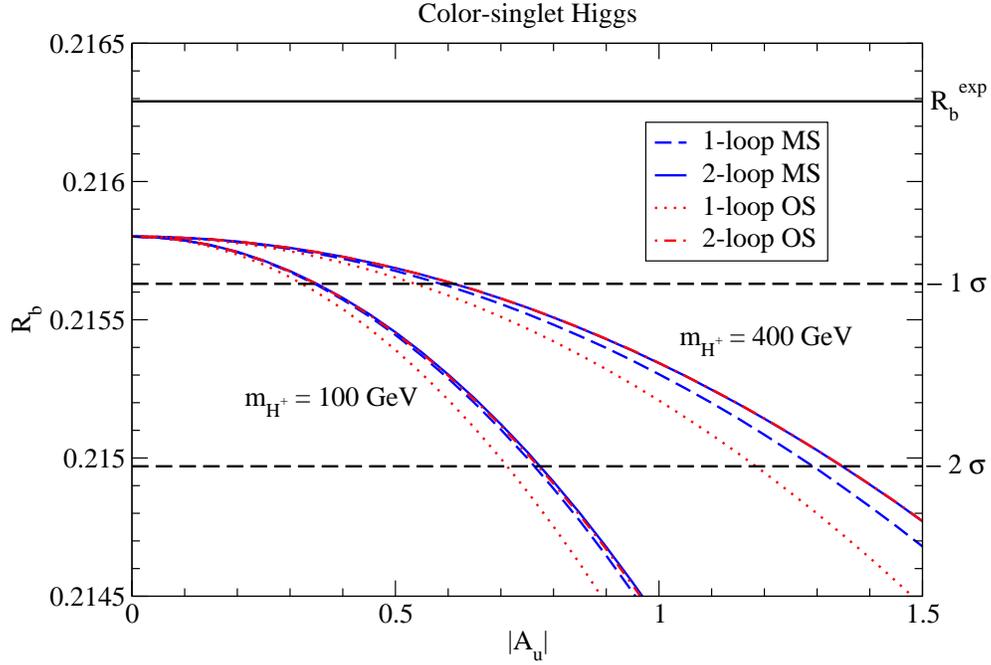}
\end{center}
\vspace{-2mm}
\caption{$R_b$ as a function of $|A_u|$ in the 2HDM with color-singlet
  Higgs, for $A_d=0$ and two different values of $\mh$. The measured
  value $R_b^{\rm exp} = 0.21629$ and the values $1\sigma$ and
  $2\sigma$ below it are displayed as horizontal lines.  For the
  meaning of the different curves see the text.}
\label{fig2}
\end{figure}

\begin{figure}[p]
\begin{center}
\epsfig{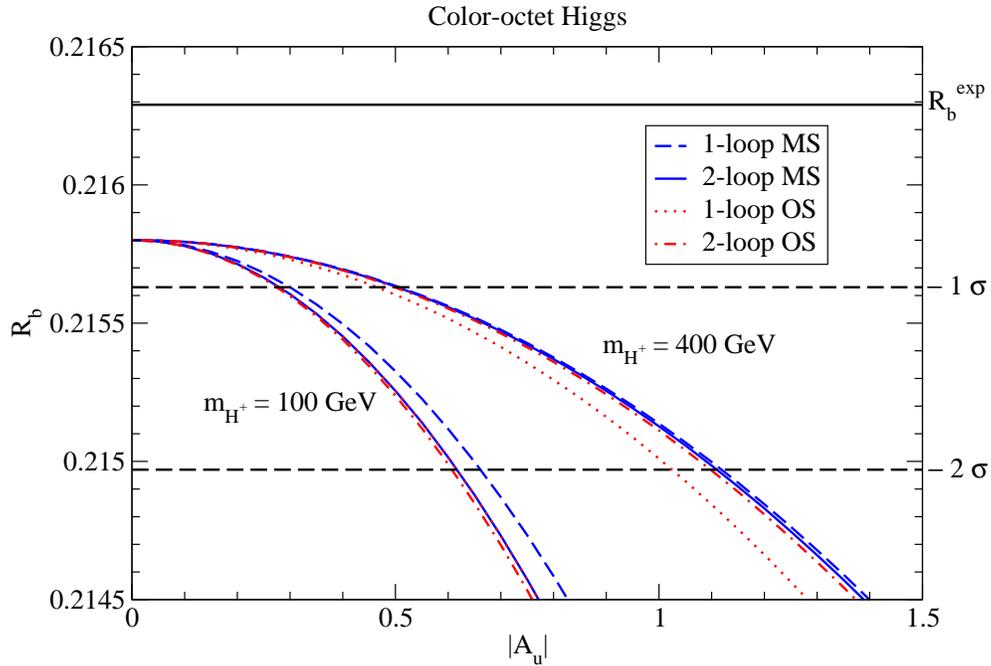}
\end{center}
\vspace{-2mm}
\caption{same as figure \ref{fig2} in the 2HDM with color-octet
  Higgs.}
\label{fig3}
\end{figure}

Figures \ref{fig2} and \ref{fig3} show our determination of $R_b$ in
the 2HDMs of type III and C, respectively, as a function of
$|A_u|$. In each plot we show two sets of curves for the charged-Higgs
mass values $\mh = 100$ GeV and $\mh = 400$ GeV. In each set, the
dashed (solid) curve represents the one-loop (two-loop) result
expressed in terms of the $\msbar$ top mass, while the dotted
(dot-dashed) curve represents the one-loop (two-loop) result expressed
in terms of the physical top mass. We also show in each plot the
measured value $R_b^{\rm exp}$ (solid horizontal line) and the values
$1\sigma$ and $2\sigma$ below (dashed horizontal lines). It can be
seen that, in both plots, the curves corresponding to the two-loop
results (with the top mass renormalized either in the $\msbar$ or in
the OS scheme) are practically overlapped. The location of the curves
corresponding to the one-loop results reflects the behavior that could
already be inferred from fig.~\ref{fig1}: in the 2HDM of type III the
one-loop result computed in terms of the $\msbar$ top mass is very
close to the two-loop result, while the one-loop result computed in
terms of the physical top mass can differ significantly. On the other
hand, in the 2HDM of type C the quality of the one-loop approximation
depends on the charged Higgs mass. At low values of $\mh$, using the
physical top mass in the one-loop result gives a much better
approximation to the two-loop result than using the $\msbar$ top mass,
while the situation is reversed at large values of $\mh$. Therefore,
only the use of the two-loop results guarantees a precise
determination of $R_b$ for all the values of $\mh$.

From figures \ref{fig2} and \ref{fig3} it is also possible to
determine the values of $|A_u|$ that are disfavored by the comparison
between $R_b^{\rm exp}$ and the corresponding theoretical prediction.
In the case of the type-III 2HDM, the two-loop curves cross the
$2\sigma$ horizontal line at $|A_u| = 0.78$ for $\mh = 100$ GeV and at
$|A_u| = 1.35$ for $\mh = 400$ GeV. In the case of the type-C 2HDM,
they cross it at $|A_u| = 0.62$ for $\mh = 100$ GeV and at $|A_u| =
1.10$ for $\mh = 400$ GeV. We checked that the crossing points for
intermediate values of $\mh$ can be determined by linear interpolation
of the values given above.

The predictions of $R_b$ for the type-II 2HDM (where $A_d=-1/A_u =
-\tan\beta$) are virtually indistinguishable from those presented in
figure \ref{fig2} for the type-III 2HDM as soon as $|A_u|>0.1$. The
upper bounds on $|A_u|$ discussed above translate, for the type-II
2HDM, into $|A_d| > 1.28$ for $\mh = 100$ GeV and $|A_d| > 0.74$ for
$\mh = 400$ GeV.  On the other hand, when $|A_u|$ tends to zero the
predictions of $R_b$ in the type-II 2HDM decrease quickly, and get two
standard deviations below $R_b^{\rm exp}$ for $|A_u| \approx 0.01$,
corresponding to $|A_d| \approx 100$. As we will see in the next
section, the bounds on $A_d$ coming from the process $\Bsg$ can be much
stronger than that, but they do not apply to the type-II 2HDM.


\section{Charged-Higgs contribution to $\Bsg$ at the NLO}

The branching ratio for $\Bsg$ is fully known at the NLO for a 2HDM of
types I--III. To cover also the case of a type-C 2HDM, the only
missing ingredient is the determination of the colored-scalar ${\cal
  O}(\as)$ contribution to the Wilson coefficients. We compute it
following the analogous computation for the type I--II 2HDM presented
in ref.~\cite{CDGG1}.
 
In the operator basis defined in ref.~\cite{CDGG1} we write the Wilson
coefficients at the scale $\mu_W$, where the ``full" theory is matched
to an effective theory with five quark flavors, as
\beq
C^{}_i(\muw) =  C^{(0)}_i(\muw) + \delta C_{i}^{(0)}(\muw)
 + \frac{\as(\muw)}{4 \pi} \left[ C^{(1)}_i(\muw) +
\delta C_{i}^{(1)}(\muw) \right]~,
\label{wc1}
\eeq
where $C^{(k)}_i(\muw)$ represents the SM contribution ($k=0,1$) while
$ \delta C_{i}^{(k)}(\muw)$ represents the charged-Higgs
contribution. At the LO, the latter is given by
\bea
\delta C_{i}^{(0)}(\muw) &= &0 ~~~~~i=1,...,6~,\\
\delta C_{7}^{(0)}(\muw)  &= & C^1_R \left( \frac{|A_u|^2}3 
F_{7}^{(1)}(y) - A_d  A_u^*  F_{7}^{(2)}(y) \right)~,
\label{eqc7} \\
\delta C_{8}^{(0)}(\muw)  &= & C^1_R  \left( \frac{|A_u|^2}3
F_{8}^{(1)}(y) -A_d A_u^*  F_{8}^{(2)}(y) \right) + C^2_R \left( 
|A_u|^2 F_{8}^{(3)}(y) + A_d A_u^*  F_{8}^{(4)}(y)\right)~,
\label{eqc8}  
\eea
where
\beq
F_7^{(1)}(y)=\frac{y(7-5y-8y^2)}{24(y-1)^3}+\frac{y^2(3y-2)}{4(y-1)^4}\ln y,
~~~~~~~~F_7^{(2)}(y)= \frac{y(3-5y)}{12(y-1)^2}+\frac{y(3y-2)}{6(y-1)^3}\ln y,
\eeq
\beq
F_8^{(1)}(y)= \frac{y(2+5y-y^2)}{8 (y-1)^3}-\frac{3y^2}{4(y-1)^4}\ln y,
~~~~~~~~F_8^{(2)}(y)=\frac{y(3-y)}{4(y-1)^2}-\frac{y}{2(y-1)^3}\ln y,
\eeq
\beq
F_8^{(3)}(y)=\frac{y(1+y)}{16(y-1)^2}-\frac{y^2}{8(y-1)^3}\ln y,
~~~~~~~~F_8^{(4)}(y)=-\frac{y}{4(y-1)}+\frac{y}{4(y-1)^2}\ln y,
\eeq
with
\beq
y=\frac{\hat m_t^2(\muw )}{\mh^2}~,
\eeq
expressed in terms of the NLO top-quark running mass at the scale
$\muw$ and of the OS charged-Higgs mass. Again, $C^1_R=1, C^2_R = 0
\:[C^1_R=C_F, C^2_R = C_A =N_c]$ for Higgs fields in the $({\bf
  1},{\bf 2})_{1/2}\: [({\bf 8},{\bf 2})_{1/2}]$ representation.

At the NLO, the charged-Higgs contributions to the Wilson coefficients
are 
\bea 
\delta C_{i}^{(1)}(\muw) & = & 0 ~~~~~i=1,2,3,5,6 ~,\\ 
\delta C_{4}^{(1)}(\muw) & = & E^H(y)~,\\ 
\delta C_7^{(1)}(\muw) &=& G_7^H(y) + \Delta_7^H(y) \ln\frac{\muw^2}{\mh^2}~,
\label{c7NLO} \\ 
\delta C_8^{(1)}(\muw) &=& G_8^H(y) + \Delta_8^H(y) \ln\frac{\muw^2}{\mh^2}~.
\label{c8NLO}
\eea 

The expressions for $G^H_{7,8}, \:\Delta^H_{7,8}$ and $E^H$ are rather
long and they are reported in the appendix.  As expected, the $\muw$
dependence in $\delta C_{7,8}$ cancels out at ${\cal O}( \as)$ because
the functions $\Delta^H_{7,8}$ entering eqs.~(\ref{c7NLO}) and
(\ref{c8NLO}) satisfy the relation
\beq 
\Delta^H_i=\gamma_0^m y \frac{\partial\, \delta C_i^{(0)}}{\partial
  y}+\frac{1}{2} \sum_{j=1}^8\gamma_{ji}^{(0)eff} \delta C_j^{(0)}~,
\label{canc}
\eeq
where $\gamma_0^m = 8$ is the LO anomalous dimension of the top mass,
while $\gamma_{ji}^{(0)eff}$ is the matrix of LO anomalous dimensions
of the Wilson coefficients, whose entries can be found in eq.~(8) of
ref.~\cite{CMM}.  
\begin{figure}[t]
\begin{center}
\epsfig{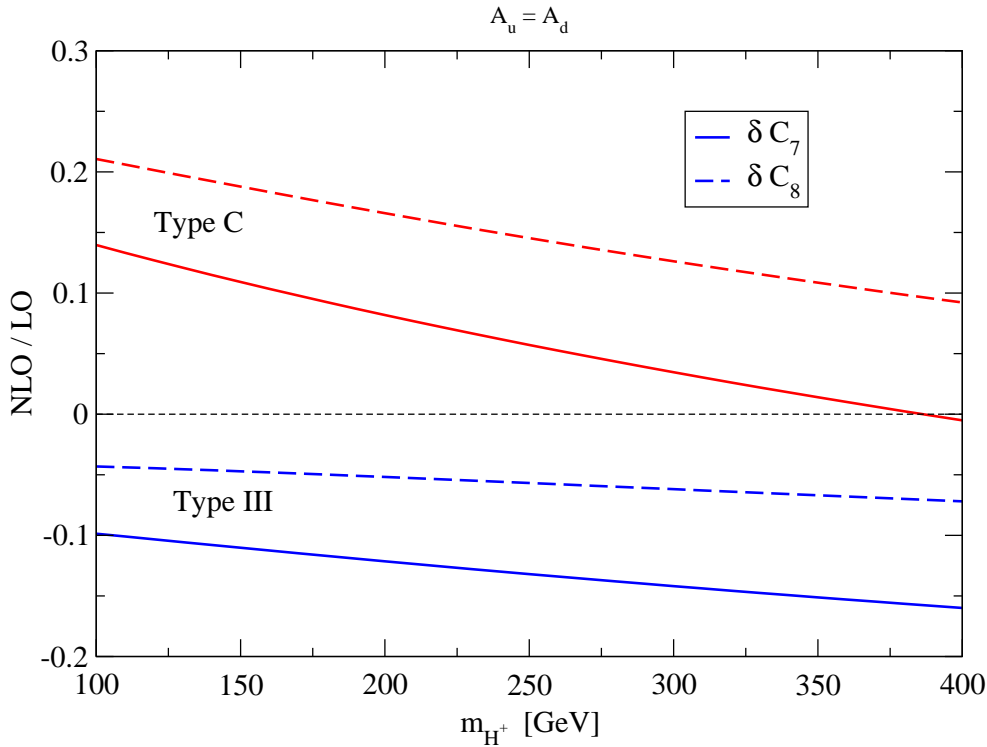}
\end{center}
\vspace{-2mm}
\caption{the ratio of NLO to LO charged-Higgs contributions to the
  Wilson coefficients $C_7$ (solid line) and $C_8$ (dashed line) as a
  function of $\mh$, with $A_d = A_u$, for the model with color-octet
  Higgs (type C) or color-singlet Higgs (type III).}
\label{fig3bis}
\end{figure}

In fig.~\ref{fig3bis} we show the ratio between the NLO and LO
charged-Higgs contributions to the Wilson coefficients $C_{7,8}$ as a
function of $\mh$, for both the type-C (color-octet) and type-III
(color-singlet) cases, with the particular choice $A_d=A_u$ (in the
color-singlet case this coincides with the type-I 2HDM). For the
type-C 2HDM, the NLO corrections can reach up to $\sim 20\%$ of the LO
contribution at small $\mh$, and they decrease as the charged-Higgs
mass increases, eventually crossing zero. In contrast, for the
type-III 2HDM the two-loop corrections to the Wilson coefficients are
always different from zero, and have opposite sign with respect to the
LO contributions. We checked that the behavior of the NLO corrections
is qualitatively similar to the one described above even when we allow
$A_d$ to take on values different from $A_u$.

The calculation of $\bratio$ is performed using a modified version of
the fortran code {\tt SusyBSG} \cite{SBSG}. The code provides a NLO
evaluation of the branching ratio in the MSSM with MFV, including the
full two-loop gluino contributions to the Wilson coefficients
\cite{DGS}. The current public version (1.3) includes also the options
of evaluating the branching ratio in the SM, in the type-II 2HDM and
in the MSSM with two-loop gluino contributions computed in the
effective Lagrangian approximation. We enlarged the 2HDM option to
include also the type-I, type-III and type-C models, thus covering all
four types of 2HDM compatible with MFV.\footnote{A public version of
  {\tt SusyBSG} with this new feature will be released soon.} The
relation between the Wilson coefficients and $\bratio$ is computed at
NLO along the lines of ref.~\cite{GM}, but the free renormalization
scales entering the NLO calculation are adjusted in such a way as to
mimic the Next-to-Next-to-Leading Order (NNLO) contributions presented
in ref.~\cite{Mal}. When the SM input parameters are set to the
partially outdated values used in ref.~\cite{Mal}, {\tt SusyBSG} gives
a SM prediction for $\bratio$ of $3.15 \times 10^{-4}$, in full
agreement with the NNLO result of that paper. Very good numerical
agreement is also found with the results of the partial NNLO
implementation of the type-II 2HDM in ref.~\cite{Mal}, which combines
NNLO anomalous dimensions and matrix elements with NLO Wilson
coefficients. We take into account a recent update \cite{GG} in the
calculation of the normalization factor for the branching ratio as
well as the latest central value of the top mass \cite{top}, which
results in a modest enhancement of the SM prediction for $\bratio$ to
$3.28 \times 10^{-4}$.

\begin{figure}[t]
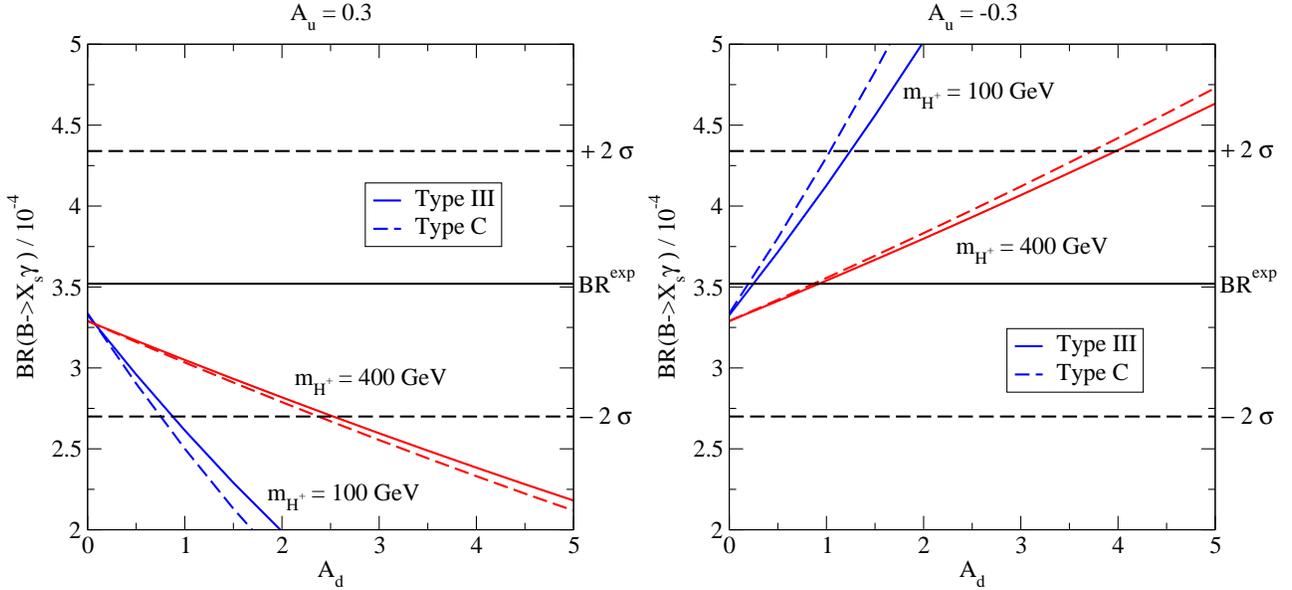

\begin{center}
\epsfig{figure=BR_vs_Ad_same.eps,width=8.4cm}
\epsfig{figure=BR_vs_Ad_oppo.eps,width=8.4cm}
\end{center}
\vspace{-1cm}
\caption{$\bratio$ as a function of $A_d$ in the type-III (solid line)
  and type-C (dashed line) models for $A_u=0.3$ (left panel) and
  $A_u=-0.3$ (right panel) and two different values of $\mh$. The
  horizontal dashed lines specify the $2 \sigma$ interval around the
  experimental value of $\bratio$.}
\label{fig5}
\end{figure}

In the 2HDMs of types I and II, the requirement of consistency between
the theoretical prediction and the measured value of $\bratio$ allows
us to set bounds on the parameters $\mh$ and $\tan\beta$, the latter
determining both Higgs-quark couplings $A_u$ and $A_d$. More
specifically, in the type-I 2HDM the charged-Higgs contribution to the
Wilson coefficients $C_{7,8}$ scales like $1/\tan\beta^2$, therefore
it is possible to derive, for each given value of $\mh$, a lower bound
on $\tan\beta$ (i.e., an upper bound on $A_u=A_d$) which is however
much less stringent than the corresponding bound derived from
$R_b$. In the type-II 2HDM there is a $\tan\beta$-independent
contribution to the Wilson coefficient, which allows to set an
absolute lower bound on $\mh$. These bounds have been extensively
discussed in the literature (see, e.g.,
refs.~\cite{Gfit,CKMfit,UTfit}) and we will not further consider them
here.
In the models of type III and C, on the other hand, the parameters
$A_u$ and $A_d$ are unrelated to each other. As can be seen in
eqs.~(\ref{eqc7}) and (\ref{eqc8}), the charged-Higgs contributions to
the Wilson coefficients $C_{7,8}$ include two terms controlled by
$|A_u|^2$ and $A_d A_u^*$, respectively.  The bounds on $A_u$ derived
in the previous section tell us that the former term cannot be too
large, while the latter can be significant for large values of
$A_d$. Furthermore, its effect on the branching ratio depends on the
relative sign between $A_u$ and $A_d$.

In fig.~\ref{fig5} we show $\bratio$ as a function of $A_d$, for both
the type-III (solid lines) and type-C (dashed lines) 2HDM, and for the
representative choices $A_u = \pm 0.3$ (the latter are allowed by
$R_b$, as can be seen in figs.~\ref{fig2} and \ref{fig3}). The left
panel displays the case of same sign between $A_u$ and $A_d$, while
the right panel shows the case of opposite sign. Each plot contains
two sets of curves corresponding to $\mh = $ 100 GeV and $\mh = 400$
GeV, respectively. The horizontal dashed lines mark the 95\% C.L.~band
around the experimental value $\bratio = (3.52 \pm 0.25)\times
10^{-4}$ \cite{HFAG}. The band also includes, added in quadrature, the
theoretical error on the 2HDM prediction (we conservatively estimate
this error as 10\% of the SM prediction for the branching ratio).
From fig.~\ref{fig5} it is clear that, unless $|A_u|$ is extremely
small, the process $\Bsg$ sets stringent limits on $A_d$. Focusing on
the case of color-singlet Higgs we see that, for $A_u = 0.3$, the
values of $A_d$ that allow for a branching ratio inside the 95\%
C.L.~band are $A_d \leq 0.9$ for $\mh = 100$ GeV and $A_d \leq
2.5$ for $\mh= 400$ GeV; for $A_u = -0.3$ the bounds are a little less
stringent, i.e.  $A_d \leq 1.3$ for $\mh = 100$ GeV and $A_d \leq 4$
for $\mh = 400$ GeV.
From the figure it is also apparent that the bound on $A_d$ for a
fixed value of $A_u$ is almost independent of the colored or colorless
nature of the charged Higgs, with the colored case showing only
slightly stronger bounds. However, as seen in the previous section,
the bounds on $A_u$ derived from $R_b$ are more dependent on the
nature of the Higgs, so that the allowed regions for the $A_d$
coefficient are in fact different for the color-singlet and
color-octet charged Higgs.

\begin{figure}[t]
\begin{center}
\epsfig{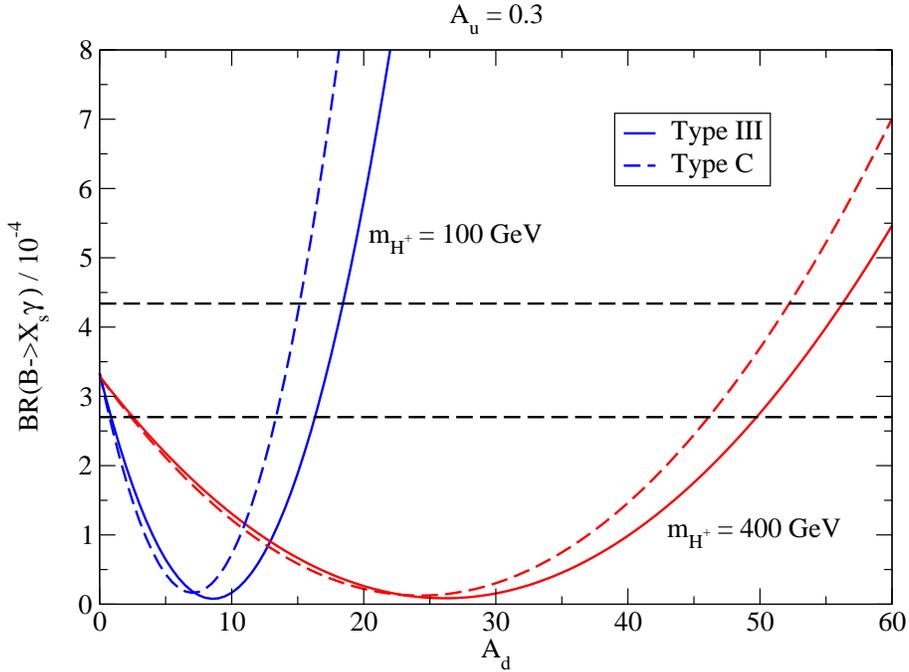}
\end{center}
\vspace{-1cm}
\caption{$\bratio$ as a function of $A_d$ in the type-III (solid line)
  and type-C (dashed line) models for $A_u=0.3$ and two different
  values of $\mh$. The horizontal dashed lines specify the $2 \sigma$
  interval around the experimental value of $\bratio$.}
\label{fig6}
\end{figure}

The case of same sign between $A_u$ and $A_d$ has the peculiarity
that, as shown in fig.~\ref{fig6}, there are actually two ranges of
values of $A_d$ that fit inside the $2\sigma$ allowed band for
$\bratio$. This is related to the fact that, in this case, the sign of
the charged-Higgs contribution to the Wilson coefficient of the
magnetic dipole operator, $C_7^H$, is opposite to the sign of the SM
contribution, $C_7^{\rm{\scriptscriptstyle SM}}$. Since $\bratio$ is
roughly proportional to $|C_7^H+C_7^{\rm{\scriptscriptstyle SM}}|^2$,
as $A_d$ increases the branching ratio goes to zero when $C_7^H\approx
-C_7^{\rm{\scriptscriptstyle SM}}$, and it goes back inside the
$2\sigma$-allowed band when $C_7^H\approx
-2\,C_7^{\rm{\scriptscriptstyle SM}}$. Thus, the two ranges of
possible values of $A_d$ differ by the sign of the amplitude ${\cal
  A}(b \to s \gamma)$, basically the sign of the Wilson coefficient
$C_7$. Although $\Bsg$ allows both ranges of values for $A_d$, there
are other observables that are sensitive to the sign of $C_7$, thus
selecting one of the two options. Among them, we cite ${\rm BR}(B \to
X_s \,l^+ l^-)$ \cite{GHM} and the isospin-breaking asymmetry that can
be constructed from the exclusive neutral and charged $ B \to K^*
\gamma$ decay modes \cite{KN,MS}. Although, for both observables,
neither the experimental result nor the theoretical prediction is at
the same level of accuracy as for $\bratio$, these observables still
give a compelling indication that the sign of ${\cal A}(b \to s
\gamma)$ is that of the SM contribution, thus eliminating the
large-$A_d$ option.


\section{Conclusions}

Minimal flavor violation is a very popular criterion that is used to
suppress FCNC effects in models with new particles at the TeV scale.
The enforcement of the MFV criterion to the simplest extension of the
SM, i.e.~a model with a second Higgs doublet, allows the possibility
of color-singlet or color-octet Higgs field. For both cases we have
considered the two-loop QCD corrections to the charged Higgs boson
contribution to $R_b$. We found that for all four types of 2HDM with
MFV, after the inclusion of the two-loop QCD corrections, the
prediction for this observable is practically insensitive to the the
choice of renormalization scheme for the top mass entering the
one-loop part of the calculation. Thus, the upper bound on the
coupling $|A_u|$ derived from $R_b$ is improved, which for the type-I
and type-II models translates in a more robust lower bound on $\tan
\beta$.  We have also computed the ${\cal O}(\as)$ contributions to
the Wilson coefficients relevant to the process $\Bsg$ for the
color-octet case.  This was the last missing ingredient to obtain a
determination of $\bratio$ at the NLO level for all 2HDM with MFV.
After the inclusion of the NLO corrections it is found that, in the
region allowed by the present experimental results, the $\Bsg$
transition is fairly insensitive to the colored or colorless nature of
the charged Higgs. Furthermore, in type-III and C models, the bounds
on the $A_d$ parameter that can be obtained from $\bratio$ and other
observables rule out the large-$A_d$ region, where effects
proportional to the bottom mass could become important.

\section*{Acknowledgments}

We thank H.~Haber for communications concerning ref.~\cite{HL}, and
P.~Gambino for useful discussions. One of us (G.D.) also thanks
M.~Nebot for his contribution in the early stage of this project.
This work was supported in part by an EU Marie-Curie Research Training
Network under contract MRTN-CT-2006-035505 (HEPTOOLS) and by ANR under
contract BLAN07-2\_194882.

\newpage


\begin{appendletterA}

  \section*{Appendix: Analytical expressions for the NLO Wilson
    coefficients}

In this appendix we report the analytic expressions for the functions
$G^H_{7,8}$ and $\Delta^H_{7,8}$ entering $\delta C^{(1)}_{7,8}$. We
find
\bea
G_7^H(y) &= &
C_R^1\,C_F \left\{ A_d A_u^*  y\left[ \frac{4(-3+7y-2y^2)}{3(y-1)^3}{\rm Li}_2
\left( 1 - \f{1}{y} \right)+
\frac{8-14y-3y^2}{3(y-1)^4}\ln^2y \right. \right.
\nonumber \\ &&~~~~~~~~~~~~~~~~~~~~~
\left. +\frac{2(-3-y+12y^2-2y^3)}{3(y-1)^4}\ln y
+\frac{7-13y+2y^2}{(y-1)^3}\right]
\nonumber \\ &&~~~~~~~~~~~
+|A_u|^2 y\left[ \frac{y(18-37y+8y^2)}{6 (y-1)^4}{\rm Li}_2
\left( 1 - \frac{1}{y} \right)+
\frac{y(-14+23y+3y^2)}{6 (y-1)^5}\ln^2y \right.
\nonumber \\ &&~~~~~~~~~~~~~~~~~~~~~
 +\frac{-50+251y-174y^2-192y^3+21y^4}{54 (y-1)^5}\ln y
\nonumber \\ &&~~~~~~~~~~~~~~~~~~~~ \left.
\left . +\frac{797-5436y+7569y^2-1202y^3}{648(y-1)^4}\right] \right\} 
\nonumber  \\
&+& C_R^1\,C_R^2 \left\{ A_d A_u^*  y\left[ \frac{-19 +25 y}{18 (y-1)^2}{\rm Li}_2
\left( 1 - \f{1}{y} \right)+
\frac{-25+33 y+15 y^2}{36 (y-1)^4}\ln^2y \right. \right.
\nonumber \\ &&~~~~~~~~~~~~~~~~~~~~~~
\left. +\frac{33-59 y+ 3 y^2}{18(y-1)^3}\ln y
+\frac{-8+31y}{36 (y-1)^2}\right]
\nonumber \\ &&~~~~~~~~~~~
+ |A_u|^2 y\left[ \frac{12-25y}{36 (y-1)^2}{\rm Li}_2
\left( 1 - \frac{1}{y} \right)+
\frac{-12+85y-108 y^2+3y^3}{72 (y-1)^5}\ln^2y \right.
\nonumber \\ &&~~~~~~~~~~~~~~~~~~~~~
 ~~~~-\frac{50+33y-195y^2+16y^3}{108 (y-1)^4}\ln y \left.
\left . +\frac{17-29y+4y^2}{18(y-1)^3}\right] \right\}~, 
\label{higceq}
\eea
\bea
\Delta_7^H(y) &= & C^1_R\,C_F \left\{ A_d A_u^* y\left[ \frac{21-47y+8y^2}{6
(y-1)^3}+\frac{-8+14y+3y^2}{3(y-1)^4}\ln y \right] \right.
\nonumber \\ && ~~~~~~~~~~~~ \left.
+|A_u|^2 y\left[ \frac{-31-18y+135y^2-14y^3}{
36(y-1)^4}+\frac{y(14-23y-3y^2)}{6(y-1)^5}\ln y \right] \right\} \nonumber \\
&+& C^1_R \,C_R^2\left\{ A_d A_u^* y\left[ \frac{1}{3 (y-1)}
-\frac{1}{3(y-1)^2}\ln y \right] \right.
\nonumber \\ && ~~~~~~~~~~~~ \left.
+|A_u|^2 y\left[- \frac{1+ y}{12(y-1)^2}+
\frac{y}{6(y-1)^3}\ln y \right] \right\}~,
\label{del7} 
\eea
\bea
G_8^H(y) &= &
C^1_R \left\{A_d A_u^* \frac13 y\left[ \frac{-36+25y-17y^2}{2(y-1)^3}{\rm Li}_2
\left( 1 - \f{1}{y} \right)+
\frac{19+17y}{(y-1)^4}\ln^2y \right. \right.
\nonumber \\ &&~~~~~~~~~~~~~~~~~~~
\left. +\frac{-3-187y+12y^2-14y^3}{4(y-1)^4}\ln y
+\frac{3(143-44y+29y^2)}{8(y-1)^3}\right]
\nonumber \\ & &~~~~~~~
+|A_u|^2 \frac16 y\left[ \frac{y(30-17y+13y^2)}{(y-1)^4}{\rm Li}_2
\left( 1 - \f{1}{y} \right)-
\frac{y(31+17y)}{(y-1)^5}\ln^2y \right.
\nonumber \\ &&~~~~~~~~~~~~~~~~~~~
 +\frac{-226+817y+1353y^2+318y^3+42y^4}{36(y-1)^5}\ln y
\nonumber \\ && ~~~~~~~~~~~~~~~~~~~ \left.
\left. +\frac{1130-18153y+7650y^2-4451y^3}{216(y-1)^4}\right] \right\}
\nonumber \\
&+& C^2_R \left\{A_d A_u^* \frac19 y\left[ \frac{-43+34y}{4(y-1)^2}{\rm Li}_2
\left( 1 - \f{1}{y} \right)+
\frac{-157-108y +81 y^2}{8(y-1)^4}\ln^2y \right. \right.
\nonumber \\ &&~~~~~~~~~~~~~~~~~~~
\left. +\frac{-51-22y+57y^2}{8(y-1)^3}\ln y
+\frac{5(13-8y)}{(y-1)^2}\right]
\nonumber \\ & &~~~~~
+|A_u|^2 \frac1{144} y\left[ \frac{-15+149y-122y^2}{(y-1)^3}{\rm Li}_2
\left( 1 - \f{1}{y} \right) \right. 
\nonumber  \\ &&~~~~~~~~~~~~~~~~~~~~ 
-\frac{15-533y - 237 y^2 + 243 y^3}{2 (y-1)^5}\ln^2y 
\nonumber \\ 
&&~~~~~~~~~~~~~~~~~~~~   
-\frac{172-744y+357y^2+23y^3}{3(y-1)^4}\ln y
 \left. \left. -\frac{203+1174y-737y^2}{2(y-1)^3}\right] \right\}
\eea
\bea
\Delta_8^H(y) &= &  C^1_R \left\{ A_d A_u^* \frac{1}{3}y
\left[ \frac{81-16y+7y^2}{2(y-1)^3}-\frac{19+17y}{(y-1)^4}\ln y \right] \right.
\nonumber \\ &&~~~~~~~   
\left. +|A_u|^2\frac{1}{6}y\left[ \frac{-38-261y+18y^2-7y^3}{
6(y-1)^4}+\frac{y(31+17y)}{(y-1)^5}\ln y \right] \right\} 
\nonumber \\
&+& C^2_R \left\{ A_d A_u^* \frac{1}{6}y\left[ \frac{31-7y}{(y-1)^2}
-\frac{19+5y}{(y-1)^3}\ln y \right] \right.
\nonumber \\ &&~~~~~~
\left. +|A_u|^2\frac{1}{12}y\left[ \frac{-19-60y+7y^2}{2(y-1)^3}
+\frac{y(31+5y)}{(y-1)^4}\ln y \right] \right\}
\label{del8} 
\eea
\bea
E^H(y)& = & C^1_R \,  |A_u|^2 \frac16 y  
\left[ \frac{16-29y+7y^2}{6(y-1)^3}+\frac{-2+3y}{(y-1)^4} \ln y \right] 
\nonumber \\
&+& C^2_R \,  |A_u|^2 \frac14 y  
\left[ \frac{-1}{(y-1)}+\frac{2+y}{3 (y-1)^2} \ln y \right]~.
\eea

The results for type I-III models are recovered setting $C^1_R =1$ and
$C^2_R =0$, while the result for the case of a colored charged scalar
in the adjoint of $SU(3)$ is obtained with $C^1_R =C_F$ and $C^2_R
=N_c$.

\vspace*{0.5cm}
\paragraph{\Large Note added:}
after the publication of our paper, we became aware of
ref.~\cite{bobeth}, which contains two-loop formulae for the Wilson
coefficients of the operators relevant to $b\rightarrow s \gamma$ in a
generic model with a heavy fermion and a heavy scalar. While
ref.~\cite{bobeth} does not specifically discuss the case of the
type-C 2HDM, the results presented in our appendix can be reproduced
with appropriate substitutions in the results of ref.~\cite{bobeth},
taking into account the different renormalization conditions on the
charged-Higgs mass. We thank Christoph Bobeth for performing this
useful comparison.

\end{appendletterA}

\end{document}